\documentclass[twocolumn,aps,prb]{revtex4}
\usepackage{graphicx}
\usepackage{amssymb}
\usepackage{amsmath}
\usepackage{bm}
\usepackage{color}
\usepackage[hidelinks=true]{hyperref}

\hypersetup{
  colorlinks   = true, %Colours links instead of ugly boxes
  urlcolor     = blue, %Colour for external hyperlinks
  linkcolor    = blue, %Colour of internal links
  citecolor   = blue %Colour of citations
}

\def\Journal #1,#2,#3,#4#5#6#7{#1 {\bf #2}, #3 (#4#5#6#7)}

\def\Vec#1{\mathbf{#1}}

\def\g1D{g_{\mathrm{1D}}}
\def\G1D{G_{\mathrm{1D}}}
\def\a1D{a_{\mathrm{1D}}}
\def\L1D{L_{\mathrm{1D}}}

\def\lsim{\, \lower -0.3ex \hbox{$<$} \kern -0.75em \lower 0.7ex \hbox{$\sim$} \,}
\def\gsim{\, \lower -0.3ex \hbox{$>$} \kern -0.75em \lower 0.7ex \hbox{$\sim$} \,}

\begin{document}

%\title{Spontaneous strain and band structure deformation in twisted bilayer graphene}
\title{Lattice relaxation and energy band modulation in twisted bilayer graphenes}

\author{Nguyen N. T. Nam$^1$ and Mikito Koshino$^2$}

\affiliation{$^1$Department of Physics, Tohoku University, Sendai 980-8578, Japan}
\affiliation{$^2$ Department of Physics, Osaka University,  Toyonaka 560-0043, Japan}

\date{\today}

\begin{abstract}
We theoretically study the lattice relaxation in the twisted bilayer graphene (TBG)
and its effect on the electronic band structure.
We develop an effective continuum theory to describe the lattice relaxation in general TBGs,
and obtain the optimized structure to minimize the total energy.
 At small rotation angles $< 2^{\circ}$, in particular, we find that the relaxed lattice drastically reduces 
the area of AA-stacking region, and form a triangular domain structure with 
alternating AB and BA stacking regions.
We then investigate the effect of the domain formation on the electronic band structure.
The most notable change from the non-relaxed model is that an energy gap up to 20meV 
opens at the superlattice subband edges on the electron and hole sides.
We also find that the lattice relaxation significantly enhances the Fermi velocity,
which was strongly suppressed in the non-relaxed model.
\end{abstract}

\maketitle

\section{Introduction}
\label{sec_introduction}

 Twisted bilayer graphene  (TBG) is a two-dimensional material where two graphene layers 
are relatively rotated by an arbitrary angle.
Such a rotational stacking structure has been widely observed in epitaxially-grown multilayer graphenes,\cite{berger2006electronic,hass2007structural,hass2008multilayer,li2009observation,miller2010structural,luican2011single, de2010epitaxial}
and also it can be fabricated by manually aligning single layers
using the transfer technique.  \cite{wang2013one, Tan2016manualTBG}
The electronic properties of TBG has also been intensively studied in theory, 
where it was shown that the energy spectrum sensitively depends on its rotation angle $\theta$.
\cite{lopes2007graphene,mele2010commensuration,trambly2010localization,shallcross2010electronic,morell2010flat,bistritzer2011moirepnas,kindermann2011local,xian2011effects,PhysRevB.86.155449,moon2013opticalabsorption,moon2012energy,bistritzer2011moireprb}
In a small $\theta$, in particular, the interference between the incommensurate lattice structures gives rise to 
a moir\'{e} pattern with a long spacial period,
and it significantly influences the low-energy spectrum.
For TBG less than a few degree, in particular,
the Dirac cones of the two layers are strongly hybridized by the moir\'{e} interlayer interaction, 
where the linear dispersion is distorted into superlattice subbands. 
\cite{lopes2007graphene,mele2010commensuration,trambly2010localization,shallcross2010electronic,morell2010flat,bistritzer2011moirepnas,kindermann2011local,xian2011effects,PhysRevB.86.155449,moon2013opticalabsorption,moon2012energy,bistritzer2011moireprb}. 
The characteristic features of superlattice band structure of TBG
were actually observed in recent experiments. \cite{kim2016charge,cao2016superlattice,kim2017tunable}

\begin{figure}
\begin{center}
 \leavevmode\includegraphics[width=0.88\hsize]{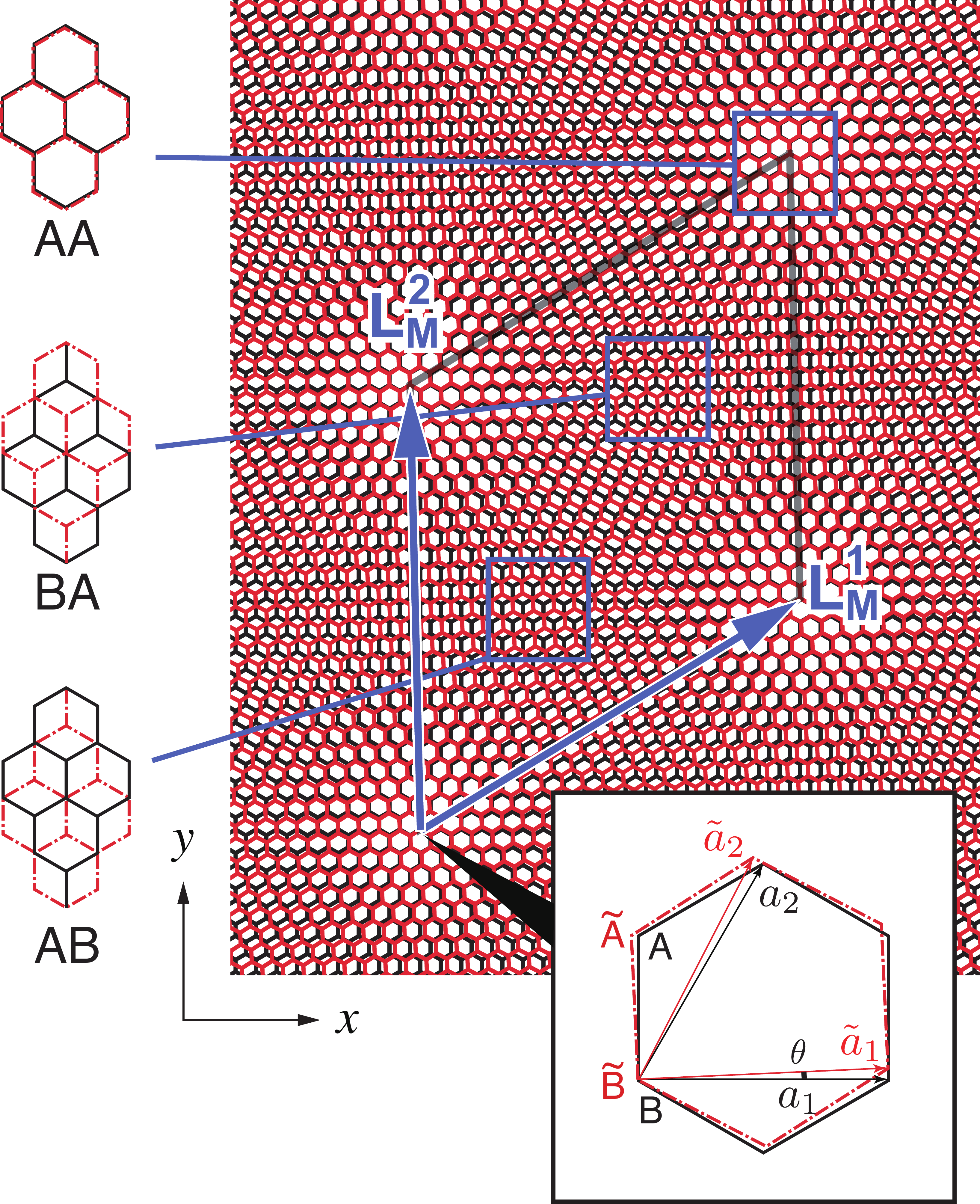}
\end{center}
\caption{Twisted bilayer graphene at rotation angle $\theta = 2.65^{\circ}$. The blue squares indicate the regions
where the lattice structure locally resembles the regular stacking arrangement such as AA, AB and BA (see the text).
Parallelogram is the moir\'{e} unit cell.
}
\label{fig:Local_structures_Moire}
\end{figure}

Most of  the band calculations for TBG assumes 
that the two graphene layers are rigid and simply stacked
without changing the original honeycomb lattices. 
In a real system, however, the lattice structure 
spontaneously relaxes to achieve an energetically favorable structure\cite{popov2011commensurate, uchida2014atomic,van2015relaxation,dai2016twisted, jain2016structure},
and it should influence the electronic spectrum.
If we consider a TBG with a small rotation angle as in Fig.\ \ref{fig:Local_structures_Moire}, for instance,
we notice that the lattice structure locally resembles the regular stacking such as AA, AB or BA, depending on the position.
Here AA represents the perfect overlapping of hexagons, 
while AB and BA are shifted configurations in which A(B) sublattice is right above B(A).
Since the interlayer binding energy is the lowest in AB and BA and the highest in AA stacking \cite{popov2011commensurate,Lebedeva2011,Gould2013},
the TBG spontaneously deforms so as to maximize the AB/BA areas while minimize AA area.
In fact, such an AB/BA domain structure was experimentally observed
in multilayer graphenes grown by chemical vapor deposition\cite{brown2012twinning,lin2013ac,alden2013strain},
and also captured in the theoretical calculations, 
\cite{popov2011commensurate, van2015relaxation,dai2016twisted, jain2016structure}
while its implication on the electronic band structure is still  unclear.
Similar lattice relaxation was also found in another moir\'{e} superlattice of graphene on hexagonal boron-nitride,
\cite{dean2013hofstadter,ponomarenko2013cloning,hunt2013massive,yu2014hierarchy,PhysRevB.89.075401,moon2014electronic} 
where the sublattice inequality in hBN results in a hexagonal domain pattern.
\cite{Woods2014GhBN, san2014spontaneous,San-Jose2014GhBN,jung2015origin}

 \begin{figure}
     \begin{center}
     \leavevmode\includegraphics[width=0.7\hsize]{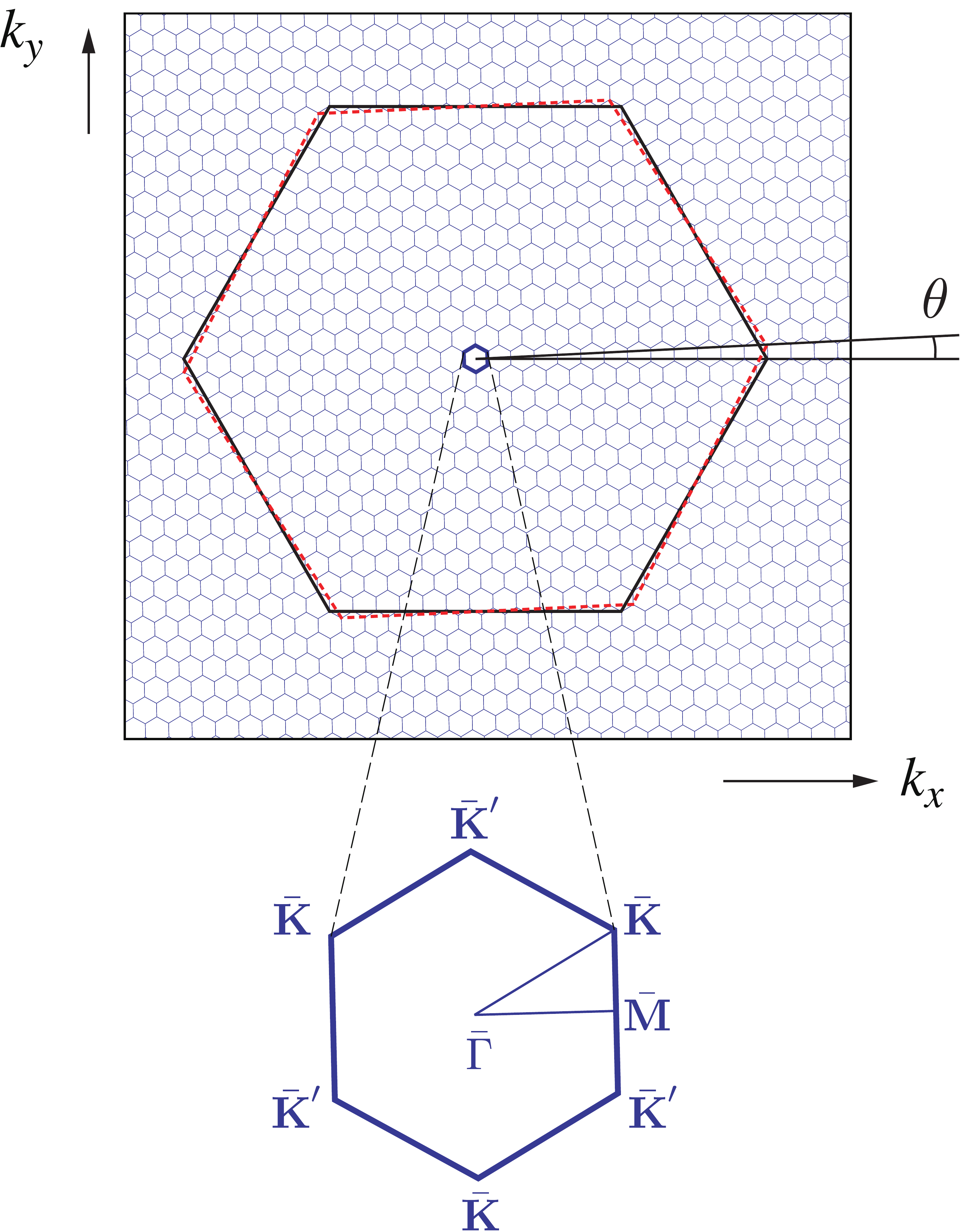}
     \end{center}
     \caption{ Brillouin zones of layer 1 (black hexagon), 
      layer 2 (red dashed hexagon) and twisted bilayer graphene (blue hexagons) at $\theta = 2.65^{\circ}$.
      }
      \label{fig:Lattice_Constant_BZ}
\end{figure}	

In this work, we present a theoretical study on the lattice relaxation in TBG
and its effect on the electronic band structure.
First, we develop a minimum phenomenological model
to describe the AB/BA domain formation in TBGs with general angles.
Using the elastic theory and a simple interlayer adhesion potential,
we express the total energy as a functional of the lattice deformation $\Vec{u}(\Vec{r})$,
and optimize it by solving the Euler-Lagrange equation.
In decreasing the rotation angle $\theta$, we actually find that 
$\Vec{u}(\Vec{r})$ increases and eventually forms a sharp domain structure 
with AB and BA regions clearly separated into a triangular pattern.
%The physics of domain formation is characterized by a single dimensionless parameter $\eta$,
%which depends on the lattice stiffness, the interlayer coupling energy, and 
%the size of the moir\'{e} unit cell.

We then investigate the effect of the domain formation on the electronic spectrum of TBG.
The lattice distortion is expected to affect the band structure 
by modifying the moir\'{e} pattern, and also by adding the strain-induced vector potential
to graphene's Dirac electron. \cite{suzuura2002phonons,pereira2009strain,guinea2010energy}
Here we calculate the band structure of relaxed TBGs by the tight-binding model
and compared it with the non-relaxed case.
A significant difference is observed in $\theta < 2^\circ$,
where the domain structure becomes pronounced.
The most notable effect is that an energy gap up to 20meV appears
at the superlattice subband edge between the lowest and the second subband, 
while it was hardly found in the non-relaxed model.
We also find that the lattice deformation significantly enhances the Fermi velocity,
which was strongly suppressed in the non-relaxed model.\cite{TramblyDeLaissardiere2012confined,Bistritzer2011magicangle}
The associated lattice distortion induces the pseudo magnetic field more than 30T,
which alternates in space with the moir\'{e} superlattice period.

The paper is organized as follows.
In Sec.\ \ref{sec_geometry}.
we introduce the lattice geometry of TBG
and the description of the moir\'{e} pattern.
In Sec.\ \ref{sec_lattice}, we develop the effective continuum theory
for the lattice relaxation.
First we consider the simpler one-dimensional model as an intuitive example,
and then we extend the model to two-dimensional TBG.
In Sec.\ \ref{sec_band}, we calculate the electronic band structure of relaxed TBGs
and discuss the effect of the lattice deformation.
A short conclusion is given in Sec.\ \ref{sec_conclusion}.
    
\section{Geometry of twisted bilayer graphene}
\label{sec_geometry}	
      		
Let us consider a TBG lattice as illustrated in Fig. \ref{fig:Local_structures_Moire}.
Here we specify the stacking geometry by starting from the AA-stacking bilayer graphene,
and rotate the layer 2 with angle $\theta$ with respect to the layer 1.
In Fig. \ref{fig:Local_structures_Moire}, we take $\theta = 2.65^\circ$ as an example.
We define the primitive lattice vectors of layer 1 as $\textbf{a}_1 = a(1,0)$,  
$\textbf{a}_2 = a(1/2,\sqrt{3}/2)$, where $a = 0.246$ nm is graphene's lattice constant.
The primitive lattice vectors of layer 2 can be obtained by rotating those of layer 1 as 
$\tilde{\textbf{a}}_i =R(\theta) \textbf{a}_i\, (i=1,2)$ where  $R(\theta)$ is rotation matrix. 
The reciprocal lattice vectors of layer 1 are given by $\Vec{a}^*_1 = 2\pi/a (1,-1/\sqrt{3})$ 
and $\Vec{a}^*_2 = 2\pi/a(0,2/\sqrt{3})$,
and those of layer 2 by  $\tilde{\Vec{a}}^*_i =R(\theta) \Vec{a}^*_i$ ($i=1,2$).    

 The Brillouin zone of layer 1 and layer 2 are shown in Fig. \ref{fig:Lattice_Constant_BZ} by 
 two large black, red dashed hexagons, respectively.
 In TBG, they are folded into reduced Brillouin zones shown by small blue hexagons. 
We label the corner points of the folded Brillouin zone by
 $\bar{K}$  and $\bar{K}'$, the midpoint of each side by $\bar M$,
 and the zone center by $\bar\Gamma$. 

When the rotation angle is small, 
the mismatch of the lattice periods of two rotated layers 
gives rise to the long-period moir\'{e} beating pattern,
of which spatial period is estimated as follows.
In the rotation from the AA stacking,
an atom on layer 2 originally located at site $\textbf{r}_0$ (right above the layer 1's atom)
is moved to the new position $\textbf{r} = R(\theta)\textbf{r}_0$. 
Then we define the interlayer atomic shift $\bm{\delta} (\textbf{r})$ as
the in-plane position of an layer 2's atom at $\Vec{r}$ measured from its counterpart on layer 1,
i.e.,
\begin{equation}
\bm{\delta} (\textbf{r}) = \textbf{r} - \textbf{r}_0 = (1 - {R}^{-1}) \textbf{r}.
\label{eq_delta}
\end{equation}
When $\bm{\delta}(\Vec{r})$ coincides with a lattice vector
of layer 1, then the position $\Vec{r}$ (layer 2's atom) is occupied 
also by an atom of layer 1, so that
the local lattice structure at $\Vec{r}$ takes AA arrangement as in the origin.
Therefore, the primitive lattice vector of the moir\'{e} superlattice $\Vec{L}_i^{\rm M}$ is 
obtained from the condition
$\bm{\delta}(\Vec{L}_i^{\rm M}) = \Vec{a}_i$, which leads to
\begin{equation}
  \Vec{L}_i^{\rm M} = (1-R^{-1})^{-1} \Vec{a}_i\quad (i=1,2).
\end{equation}
The lattice constant $L_{\rm M} = | \Vec{L}_1^{\rm M}|=| \Vec{L}_2^{\rm M}|$
is given by
 \begin{equation}
       L_{\rm M} =  \frac{a}{2\sin (\theta/2)}.
       \end{equation}
The corresponding moir\'{e} reciprocal lattice vectors 
satisfying $\Vec{G}^{\rm M}_i\cdot\Vec{L}_j^{\rm M} = 2\pi\delta_{ij}$
are written as
\begin{equation}
  \Vec{G}_i^{\rm M} = (1-R)\, \textbf{a}^*_i
  =  \textbf{a}^*_i - \tilde{\textbf{a}}^*_i. \quad (i=1,2),
\end{equation}
where we used $R^\dagger = R^{-1}$.
       
In general TBGs, the lattice structure is not exactly periodic in the atomic level, since the
moir\'{e} interference pattern is not  generally commensurate with the lattice period.
However, the superlattice becomes rigorously periodic at some special $\theta$,    
where vector   $m\textbf{a}_1 + n\textbf{a}_2$ meets vector $n\textbf{a}'_1 + m\textbf{a}'_2$
with certain integers $m$ and $n$.\cite{mele2010commensuration}
The exact superlattice period is then given by 
\begin{eqnarray}
 L = |m \Vec{a}_1 + n \Vec{a}_2| = a \sqrt{m^2 + n^2 + mn}
= \frac{|m-n|a}{2\sin(\theta/2)},
\label{eq_superlattice_period}
\end{eqnarray}
which is $|m-n|$ times as big as the moir\'{e} period $L_{\rm M}$.
The rotation angle $\theta$ is equal to the
angle between two lattice vectors $m \Vec{a}_1 + n \Vec{a}_2$ and
$n \Vec{a}_1 + m \Vec{a}_2$, and it is explicitly given by
\begin{eqnarray}
 \cos\theta = \frac{1}{2}\frac{m^2+n^2+4mn}{m^2+n^2+mn}.
\end{eqnarray}
In Table \ref{tbl_TBGs}, we present $(m,n)$, the rotation angle  $\theta$,
the size of the moir\'{e} unit cell $L_{\rm M}$, and the dimensionless parameter $\eta$ 
(introduced in Sec.\ \ref{sec_lattice})
for several TBGs considered in this paper. 

\begin{table}
$
\begin{array}{clcc}
 (m,n) & \theta[^\circ] & L_{\rm M} [{\rm nm}] & \eta \\ 
\hline
(12, 13) & 2.65 & 5.33 & 0.258 \\
(22, 23) & 1.47 & 9.59 & 0.464 \\
(27, 28) & 1.20 & 11.72 & 0.567 \\
(31, 32) & 1.05 & 13.42 & 0.650 \\ 
(33, 34) & 0.987 & 14.27 & 0.691 \\ 
(40, 41) & 0.817 &  17.26& 0.835 \\ 
(60, 61) & 0.547 &  25.78 & 1.248
\end{array}
$
\caption{Index $(m,n)$, the rotation angle $\theta$,
the size of the moir\'{e} unit cell $L_{\rm M}$, and the dimensionless parameter $\eta$ 
(see, Sec.\ \ref{sec_lattice})
for several TBGs considered in this paper. }
\label{tbl_TBGs}
\end{table}

 \section{Optimized lattice structure}	
\label{sec_lattice}

\subsection{1-D atomic chain}
		
\begin{figure}
\begin{center}
\leavevmode\includegraphics[width=1.0\hsize]{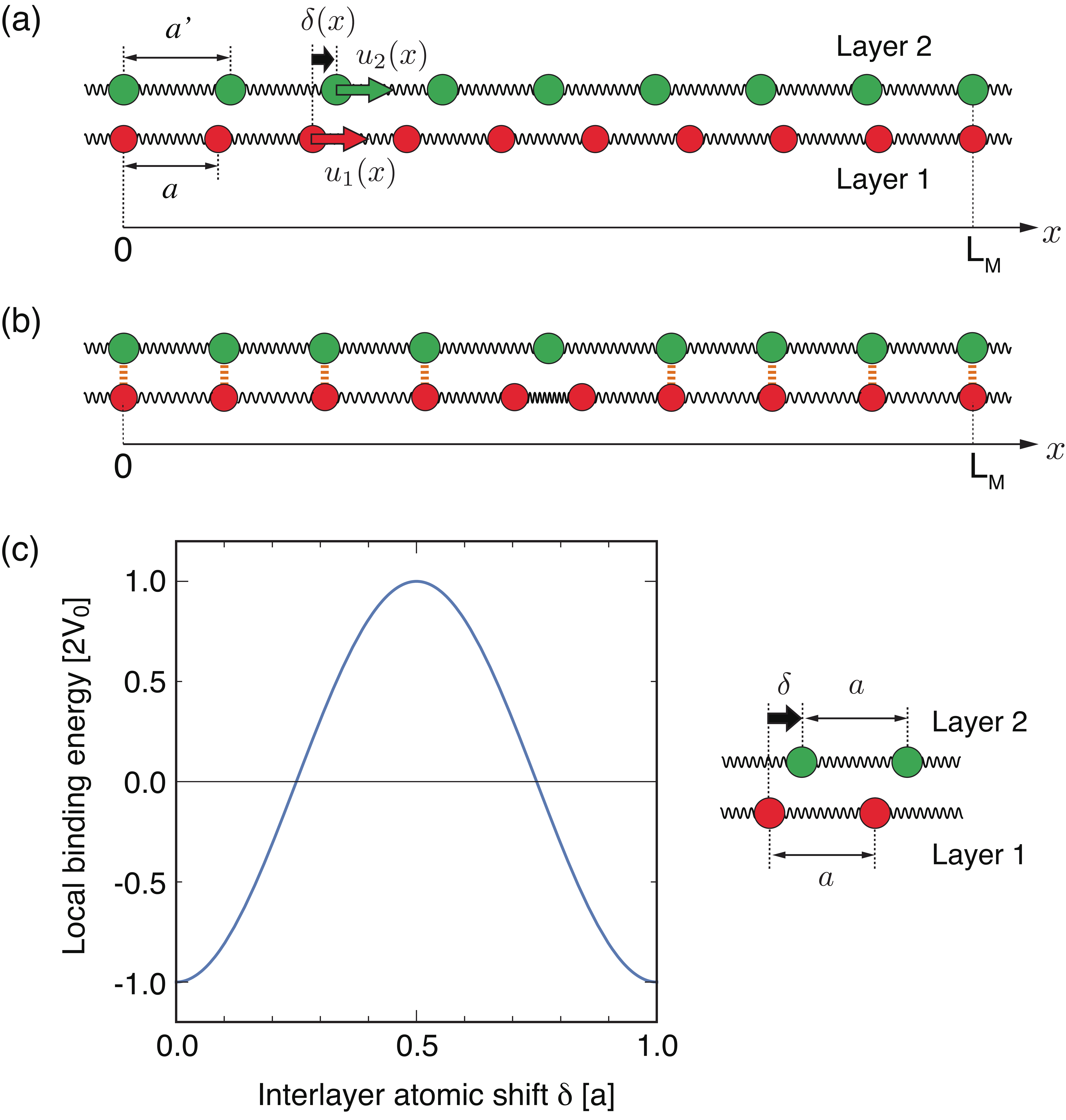}
\end{center}
\caption{(a) One-dimensional moir\'{e} superlattice model.
(b) Schematic picture of the relaxed structure, where the atoms are locked to
the vertically-aligned positions leaving a domain boundary in the middle.
(c) Interlayer binding energy per length as a function of the relative translation $\delta$
for commensurate double chains.
}
\label{fig:1D_chain}
\end{figure}	

\begin{figure}
\begin{center}
\leavevmode\includegraphics[width=0.7\hsize]{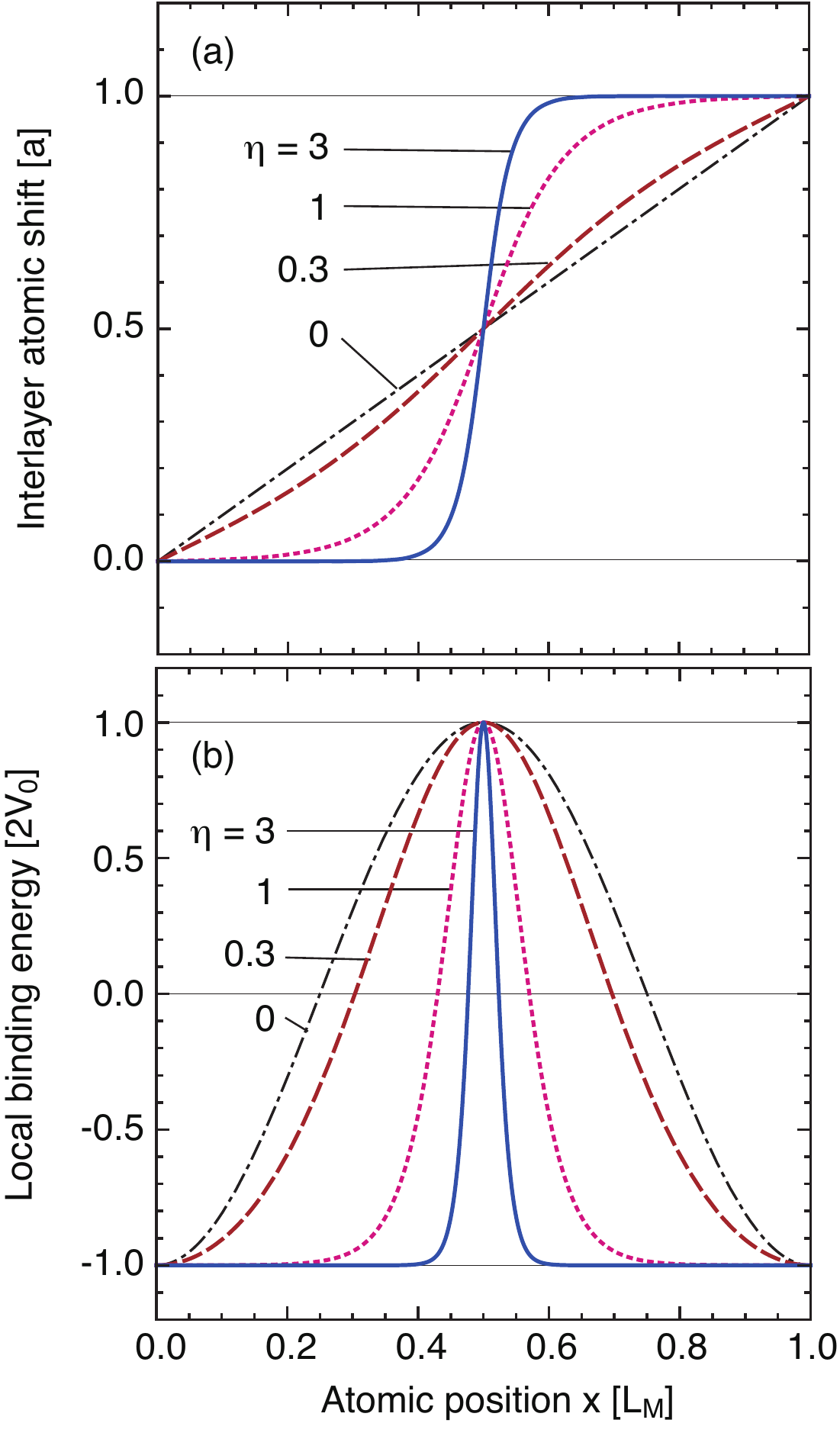}
\end{center}
\caption{
(a)  Interlayer atomic shift $\delta(x)$ and 
(b) local binding energy $V[\delta(x)]$ 
plotted against the postion $x$
in one-dimensional moir\'{e} superlattice
with $\eta =$ 0, 0.3, 1 and 3.}
\label{fig:Eta}
\end{figure}

To describe the lattice relaxation in the continuum theory, 
we start with a one-dimensional (1D) model \cite{popov2011commensurate} as a simple and intuitive example.
The extension to TBG is straightforward as we will see in the next section.
Here we consider a 1D moir\'{e} superlattice as shown in Fig.\ \ref{fig:1D_chain}(a),
which is composed of two atomic chains 1 and 2  having slightly different lattice periods,
$a = L_{\rm M}/N$ and $a'=L_{\rm M}/(N-1)$, respectively, with a large integer $N$. 
The common period of the whole system is given by $L_{\rm M}$.
Inside a supercell, there are $N$ sites and $N-1$ sites in chain 1 and 2, respectively.
This model can be viewed as an interacting two-chain version of
Frenkel-Kontorova model.

In TBG,  the atoms in different layers can be associated by rotation.
Likewise, the atoms of chain 1 and 2 are associated by expansion.
We can then define the interchain atomic shift $\delta(x)$ 
as the relative position of the site on chain 2  located at $x$
measured from the position of its counterpart on chain 1, or
\begin{equation}
\delta (x) = x - \frac{N-1}{N} x =  \frac{a}{L_{\rm M}} x.
\label{eq_delta0}
\end{equation}
This corresponds to Eq.\ (\ref{eq_delta}) in TBG.
Obviously we have $\delta (nL_{\rm M}) = na$ for integer $n$,
i.e. the atoms on different chains are vertical aligned
at $x=nL_{\rm M}$.

Now we introduce the attractive interaction between the atoms of chain 1 and 2,
while allowing the atoms move only in parallel to chain.
We expect that the atoms move their positions to reduce the interchain binding energy $U_B$.
As a result, the system tends to increase the vertically-aligned region,
and then a domain structure should be formed as schematically illustrated in Fig. \ref{fig:1D_chain}(b).
At the same time, however, such deformation increases the elastic energy $U_E$,
so the optimized state can be obtained by minimizing the total energy $U=U_E+U_B$.

Now we define $u_1(x)$ and $u_2(x)$ as displacement of atomic positions on layer 1 and layer 2, respectively,
measured from the non-relaxed state.
The interchain atomic shift in presence of deformation
is then 
\begin{align}
\delta(x) = \delta_0(x) + u_2(x) - u_1(x),
\label{eq_delta_1D}
\end{align}
where  $\delta_0(x)=(a/L_{\rm M})x $ is that in absence of the deformation, Eq.\ (\ref{eq_delta0}).
Following the standard elastic theory, we assume that the elastic energy is expressed as
\begin{equation}
\label{Eq:U_E}
U_E =  \int  \frac{1}{2} \kappa \left[ \left(\frac{\partial u_1}{\partial x} \right)^2 +\left(\frac{\partial u_2}{\partial x} \right)^2 \right]  \mathrm{d} x,
\end{equation}
where $\kappa$ is elastic constant to characterize the stiffness of the lattice.

If $a$ is very close to $a'$, the moir\'{e} superperiod $L_{\rm M}$ is much greater than the lattice constant $a$, 
and then the local lattice structure resembles commensurate chains with the identical lattice period $a$,
which are relatively shifted by some specific $\delta$ [Fig. \ref{fig:1D_chain}(c)].
Let define $V[\delta]$ as the interchain binding energy per unit length of the commensurate chains.
Here we assume an attractive interaction described by the sinusoidal function for $V[\delta]$,
\begin{align}
V[\delta] = -2V_0 \cos a^* \delta,
\label{eq_v_delta}
\end{align}
where $V_0>0$ and $a^* = 2 \pi / a$. Obviously $V[\delta]$ is periodic with period $a$,
because the sliding by the lattice spacing $a$ is equivalent to no sliding.
 It takes minimum at 
vertically aligned arrangement, $\delta= na$ ($n$: integer), and maximum at the half shift $\delta= (n+1/2)a$.
Now in the incommensurate chains in which $a$ and $a'$ are slightly different,
the interchain atomic shift $\delta$ is not a constant but slowly varying as a function of $x$.
Therefore, the interchain binding energy of the incommensurate chains as a whole is written as
\begin{equation} \label{Eq:U_B}
U_B =  \int V\left[\delta(x)\right] \mathrm{d} x.
\end{equation}
By using Eqs.\ (\ref{eq_delta0}), (\ref{eq_delta_1D}) and (\ref{eq_v_delta}), we have
$V\left[\delta(x)\right] =  -2V_0 \cos [G_{\rm M} x + a^* (u_1 - u_2)]$, where $G_{\rm M} = 2 \pi /L_{\rm M}$ is the reciprocal vector for the moir\'{e} superlattice, and we used the relation $a^*\delta_0(x) = G_{\rm M} x$.
	       	
The total energy  $U = U_B + U_E$ is the functional of $u_1(x)$ and $u_2(x)$.
Here we define the coordinates $u_\pm = u_1 \pm u_2$, and rewrite $U$
as a functional of $u_\pm$.
The optimized state to minimize the total energy is 
obtained by solving the Euler-Lagrange equations,
\begin{align}
& \kappa \frac{\partial^2 u_+}{\partial x^2} = 0,
\\
& \kappa \frac{\partial^2 u_-}{\partial x^2} -4 a^* V_0  \sin (G_{\rm M} x + a^* u_-) = 0.\label{Eq:EL}
\end{align}

In the following, we assume that the lattice deformation keeps the original superlattice period,
i.e., $u_\pm(x)$ is periodic in $x$ with period $L_{\rm M}$.
Then $u_+(x) = {\rm const.}$ is the only solution of the first equation.
To solve  the second, we apply the Fourier transformation,
 \begin{align} 
& u_-( x)  = \sum_n u_{n} e^{in G_{\rm M}x} \label{Eq:un} \\ 
& \sin (G_{\rm M}x + a^* u_- (x) ) = \sum_n f_n e^{inG_{\rm M}x}.  \label{Eq:fn}
 \end{align}
Eq.\ (\ref{Eq:EL}) is then reduced to
\begin{equation}
 u_{n} = -\frac{4a^*V_0}{\kappa(nG_{\rm M})^2} f_n.
\label{Eq:EL2}
\end{equation}
Eqs.(\ref{Eq:un}), (\ref{Eq:fn}) and  (\ref{Eq:EL2})
are a set of self-consistent equations to be solved.

% where
% \begin{align}\label{Eq:fn}
%& u_{n} = \frac{1}{L} \int_0^L u_0(x) e^{-inGx}  dx, \\ 
%& f_n = \frac{1}{L} \int_0^L dx \sin (Gx + gu_-(x)) e^{-inGx},
% \end{align}

By scaling the displacement $u_i(x)$ by $a$,
Eq.\ (\ref{Eq:EL2}) can be written in a dimensionless form,
\begin{eqnarray}
\frac{u_n}{a} = -\frac{2\eta^2}{\pi n^2} f_n, 
\label{Eq:EL3}
\end{eqnarray}
where $\eta$ is a dimensionless parameter defined by
 \begin{equation}\label{Eq:Eta}
\eta = \sqrt{\frac{V_0}{\kappa}} \frac{L_{\rm M}}{a}.
 %\eta = \frac{V_0 L_{\rm M}^2}{\kappa a^2}.
 \end{equation}
Roughly speaking, the parameter $\eta$ characterizes how many harmonics are relevant in the displacement $u_-(x)$.
Since $f_n$ is of the order of 1, 
the condition that $u_n$ is comparable to $a$ is given by $2\eta^2/(\pi n^2) \gsim 1$, or
\begin{equation}
n \lsim \sqrt{\frac{2}{\pi}}\,\eta.
%n \lsim \sqrt{\frac{2\eta}{\pi}}.
\end{equation}

When $\eta$ is small such that $(2/\pi)^{1/2}\eta\ll 1$,
only the first harmonic term is relevant so $u_-(x)$ is well approximated by a single sinusoidal function.
This situation occurs in stiff lattice (large $\kappa$), weak interchain interaction (small $V_0$)
or small moir\'{e} period (small $L_{\rm M}$).
When $\eta$ is large, on the contrary,
the large number of harmonics are significant so that $u_-(x)$ becomes a sharp function 
with respect to the moir\'{e} period $L_{\rm M}$.
This condition corresponds to soft lattice, strong interaction, and large moir\'{e} period.

The self-consistent equations Eqs.(\ref{Eq:un}), (\ref{Eq:fn}) and  (\ref{Eq:EL2})
can be solved by numerical iteration
with higher harmonics appropriately truncated.
In Fig. \ref{fig:Eta}(a), we plot the interlayer atomic shift $\delta(x)$ 
for the optimized state at some different $\eta$'s. 
The line of $\eta = 0$ represents non-relaxed case $\delta_0(x) = ax/L$, 
and the relative shift from this line represents the displacement $u_-(x)$.
The actual displacement on each chain is  given by $u_1 = -u_-/2$ and $u_2=u_-/2$.
At $\eta = 0.3$, $u_-$ is small compared to the atomic spacing $a$, 
and it contains only low frequency Fourier components.
In increasing $\eta$, the $u_-$ becomes larger and at the same time higher harmonics become more relevant.
In $\eta = 3.$, we clearly see a step-like structure
consisting of two plateau regions of $\delta = 0$ and $a$,
which are nothing but domains where the atoms are locked to
the vertically-aligned positions. 
Fig.\ \ref{fig:Eta}(b) presents the corresponding plot for
the local binding energy $V[\delta(x)]$.
We see that the original cosine function at $\eta=0$ is gradually deformed
so as to expand the plateau regions.
In $\eta = 3$, the system achieves the minimum energy almost everywhere,
except for a thin domain boundary in the middle.
 
Actually, the sharp domain boundary observed in large $\eta$ is well approximated by an analytical soliton 
solution.
If we concentrate on a small region near the domain boundary centered at
$x = L_{\rm M}/2$, Eq.\ (\ref{Eq:EL}) is reduced to 
\begin{align}
\kappa \frac{\partial^2 u_-}{\partial x^{'2}} + 4 a^* V_0  \sin (a^* u_-) = 0, \label{Eq:EL_reduced}
\end{align}	
where $x' = x - L_{\rm M}/2$, and the term $G_{\rm M}x$ is approximated by $G_{\rm M}(L_{\rm M}/2) = \pi$,
assuming the domain boundary is much narrower compared to $L_{\rm M}$.
This has an exact solution \cite{popov2011commensurate} 
\begin{align}
u_-(x') = 
\frac{2a}{\pi}\arctan
\left[
\exp
\left(4\pi
\sqrt{
\frac{V_0}{\kappa}
}
\frac{x'}{a}
\right) 
\right] - \frac{a}{2},
\end{align}	
which is found to nicely agree with the numerically-obtained $u_-$ near the boundary.
Therefore, the typical width of the domain boundary is characterized by
\begin{align}
w_{\rm d} \approx \frac{a}{4}\sqrt{\frac{\kappa}{V_0}}.% =  \frac{L_{\rm M}}{4\sqrt{\eta}}.
\label{eq_domain_width}
\end{align}
Using Eqs.\ (\ref{Eq:Eta}) and (\ref{eq_domain_width}), we have
\begin{align}
\frac{w_{\rm d}}{L_{\rm M}} =  \frac{1}{4\eta},
\end{align}
so the parameter $\eta$ characterizes the ratio of the domain wall width to the moir\'{e}
unit cell.

%%%%%
\subsection{Twisted bilayer graphene}

\begin{figure}
\begin{center}
\leavevmode\includegraphics[width=1.\hsize]{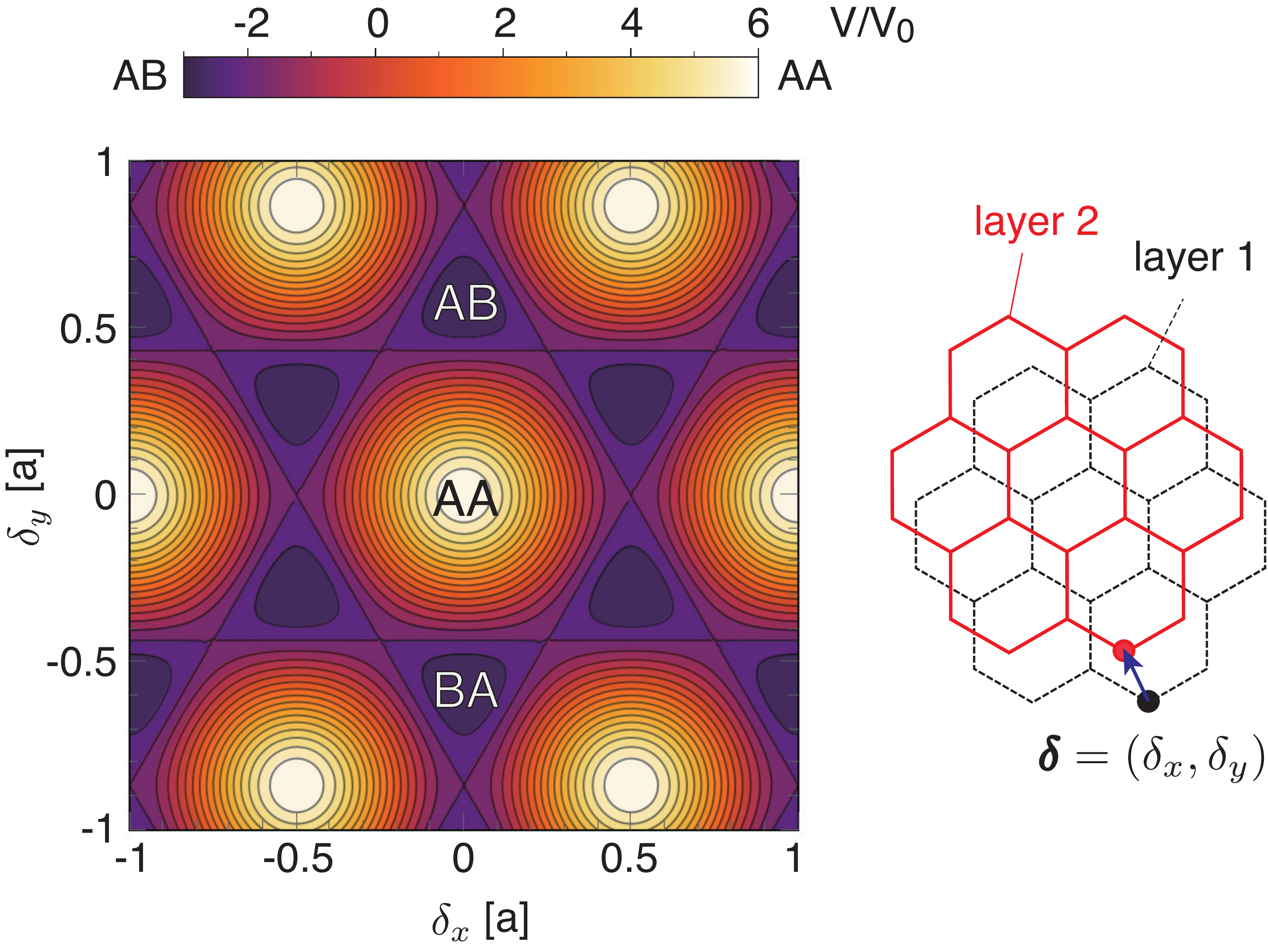}
\caption{Inter-layer binding energy $V[\bm{\delta}]$ of TBG as a function of local atomic shift $\bm{\delta} (\textbf{r})$.}
\label{fig:Binding_Energy}
\end{center}
\end{figure}

The above formulation for 1D moir\'{e} superlattice can be 
extend to TBG system in a straightforward manner.
Let us consider a TBG with a long-period moir\'{e} pattern as illustrated in Fig.\  \ref{fig:Local_structures_Moire},
and introduce the lattice deformation which is specified by the displacement vector
$\textbf{u}^{(l)}(\textbf{r})$ for layer $l =1,2$.
The interlayer atomic shift under the deformation is then given by
 \begin{equation}
\bm{\delta}(\textbf{r}) = \bm{\delta}_0 (\textbf{r}) + \textbf{u}^{(2)}(\textbf{r}) -  \textbf{u}^{(1)}(\textbf{r}),
\label{eq_delta_TBG}
 \end{equation}
where  $\bm{\delta}_0 (\textbf{r})$ is one without lattice deformation,
which is defined by Eq.\ (\ref{eq_delta}).
Here we neglect the out-of plane component of the displacement vectors
and concentrate on the in-plane motion, in order to describe the domain formation within the simplest framework.
The expected effect of the out-of-plane corrugation will be discussed later.
The elastic energy is expressed by \cite{suzuura2002phonons,San-Jose2014GhBN}
         \begin{multline}\label{Eq:U_E_TBG}
         U_E = \sum_{l=1}^2 \int \frac{1}{2} \left\{ (\lambda +\mu ) (u_{xx}^{(l)} + u_{yy}^{(l)})^2 \right. \\
       \left.   +\mu \left[ ( u_{xx}^{(l)} - u_{yy}^{(l)})^2 + 4(u_{xy}^{(l)})^2 \right] \right\} d^2 \textbf{r}
         \end{multline}
 %\begin{eqnarray}\label{Eq:U_E_TBG}
% U_E =\sum_{l=1}^2 \int \limits_{S_M} \frac{1}{2} 
 %\left[ 2\mu \text{Tr} (\textbf{u}_l^2) +\lambda (\text{Tr} (\textbf{u}_l))^2 \right] \text{d}^2 \textbf{r},		
 %\end{eqnarray}
 where $S_{\rm M}= (\sqrt{3}/2)L_{\rm M}^2$ is the area of moir\'e unit cell, 
 $\lambda \approx 3.5$ eV/$\text{\AA}^2$ and
 $\mu \approx 7.8$ eV/$\text{\AA}^2$ are typical values graphene's $\text{Lam}\acute{\text{e}}$ 
 factor \cite{san2014spontaneous}, 
and $u^{(l)}_{ij} = (\partial_i u_j^{(l)} + \partial_j u_i^{(l)})/2$ is  strain tensor.

When the moir\'{e} superperiod $L_{\rm M}$ is much greater than the lattice constant $a$, 
the local lattice structure resembles non-rotated bilayer graphene relatively shifted by $\bm{\delta}$
depending on the position [Fig.\ \ref{fig:Binding_Energy}].
We define as $V[\bm{\delta}]$ the interlayer binding energy per area of non-rotated bilayer graphene.
In the simplest approximation, it can be written as a cosine function $\bm{\delta} (\textbf{r})$ as
\begin{equation}
V[\bm{\delta}]  = \sum^3_{j= 1} 2 V_0 \cos [\textbf{a}^*_j \cdot \bm{\delta}],
\label{eq_v_TBG}
\end{equation}
where $\textbf{a}^*_3 = -\textbf{a}^*_1 - \textbf{a}^*_2$.
The function takes the maximum value $6V_0$ at AA stacking ($\bm{\delta}=0$)
and the minimum value $-3V_0$ at AB and BA stacking.
The difference between the binding energies of AA and AB/BA structure 
is $9V_0$ per area, and this amounts to $\Delta  \epsilon = 9V_0 S_G/ 4$ per atom
where $S_G$ is the area of graphene's unit cell.
In the following calculation, we use $\Delta \epsilon = 0.0189$ (eV/atom) as a typical value \cite{Lebedeva2011,popov2011commensurate}.
The potential profile of $V[\bm{\delta}]$ is presented in Fig. \ref{fig:Binding_Energy}.

In TBG, $\bm{\delta}$ is not a constant but slowly varying as a function of the 2D position.
Then the inter-layer binding energy of TBG as a whole is written as
\begin{eqnarray}\label{Eq:U_B_TBG}
U_B  &=& \int V[\bm{\delta} (\textbf{r})]   \text{d}^2 \textbf{r}.		
\end{eqnarray}
$V[\bm{\delta}]$ is periodic in $\bm{\delta}$ with the lattice period of graphene.
 By using Eqs.\ (\ref{eq_delta_TBG})	 and (\ref{eq_v_TBG}), we have
\begin{equation}
V[\bm{\delta} (\textbf{r})]   = \sum^3_{j= 1}  
2V_0 \cos [ \textbf{G}_j^{\rm M} \cdot \textbf{r} + \textbf{a}^*_j(\textbf{u}^{(2)} -\textbf{u}^{(1)}) ],
\end{equation}
where $ \textbf{G}^{\rm M}_3 = - \textbf{G}^{\rm M}_1 - \textbf{G}^{\rm M}_2$ and
 we used the relation $\textbf{a}^*_j \cdot \bm{\delta}_0(\textbf{r})=\textbf{G}_j^{\rm M} \cdot \textbf{r}$.

The relaxed state can be obtained by minimizing  total energy  $U = U_E + U_B$
as a functional of $\textbf{u}^{(l)}(\textbf{r})$.
We define $\textbf{u}^{\pm} = \textbf{u}^{(2)} \pm \textbf{u}^{(1)}$ and rewrite $U$ as a functional of $\textbf{u}^{\pm}$.
In the following we consider only  $\textbf{u}^{-}$
 as we are interested in relative displacement between atoms on two layers.
 The Euler-Lagrange equations for  $\textbf{u}^{-}$ read
\begin{multline}\label{Eq:Euler1}
\frac{1}{2} (\lambda +\mu) \left( \frac{\partial^2u^{-}_x}{\partial x^2} +\frac{\partial^2 u^{-}_y}{\partial x \partial y} \right) 
	+ \mu \left( \frac{\partial^2u^{-}_x}{\partial x^2} 
	+\frac{\partial^2 u^{-}_x}{\partial y^2} \right)\\
	 + \sum_{j=1}^3 2 V_0 \sin [\textbf{G}_j^{\rm M} \cdot \textbf{r} + \textbf{a}^*_j \cdot \textbf{u}^{-}] g^x_j = 0,
\end{multline}
\begin{multline}\label{Eq:Euler2}
\frac{1}{2} (\lambda +\mu) \left( \frac{\partial^2u^{-}_y}{\partial y^2} +\frac{\partial^2 u^{-}_x}{\partial x \partial y} \right) 
	+ \mu \left( \frac{\partial^2u^{-}_y}{\partial x^2} 
	+\frac{\partial^2 u^{-}_y}{\partial y^2} \right)\\
	 + \sum_{j=1}^3 2 V_0 \sin [\textbf{G}_j^{\rm M} \cdot \textbf{r} + \textbf{a}^*_j \cdot \textbf{u}^{-}] g^y_j = 0,
\end{multline}

We define the Fourier components  $\textbf{u}^{-}_{\textbf{q}}$ and $f^j_{\textbf{q}}(j=1,2,3)$  as
 \begin{align}
&	     \textbf{u}^{-} (\textbf{r})  = \sum \limits_{\textbf{q}} \textbf{u}^{-}_{\textbf{q}} e^{i\textbf{q}\cdot \textbf{r}},
\label{Eq:Fourier}          \\
 &          \sin \left[\textbf{G}_j^{\rm M} \cdot \textbf{r} + \textbf{a}^*_j \cdot \textbf{u}^{-}(\textbf{r})\right] = \sum_{\textbf{q}} f^j_{\textbf{q}} e^{i \textbf{q} \cdot \textbf{r}}, 
 \label{Eq:FqFunc}
\end{align}
where $\textbf{q} = m \textbf{G}^{\rm M}_1 + n \textbf{G}^{\rm M}_2$ are vectors of reciprocal superlattice.
Euler-Lagrange equations (\ref{Eq:Euler1}) and (\ref{Eq:Euler2}) are rewritten in a matrix form as 
 \begin{align} 
&        \textbf{u}^{-}_{\textbf{q}} =  \sum_{j =1}^3  4V_0 f^j_{\textbf{q}} \hat{K}_{\textbf{q}}^{-1} \textbf{a}^*_j,
\notag\\
&        \hat{K}_{\textbf{q}}=\left( {\begin{array}{cc}
        (\lambda+2\mu) q_x^2 + \mu q_y^2 &  (\lambda + \mu)q_x q_y \\
        (\lambda +\mu) q_x q_y & (\lambda + 2\mu)q_y^2 + \mu q_x^2
        \end{array}} \right) .
        \label{Eq:E-LMatrix}
 \end{align}

\begin{figure}
\begin{center}
\leavevmode\includegraphics[width=0.7\hsize]{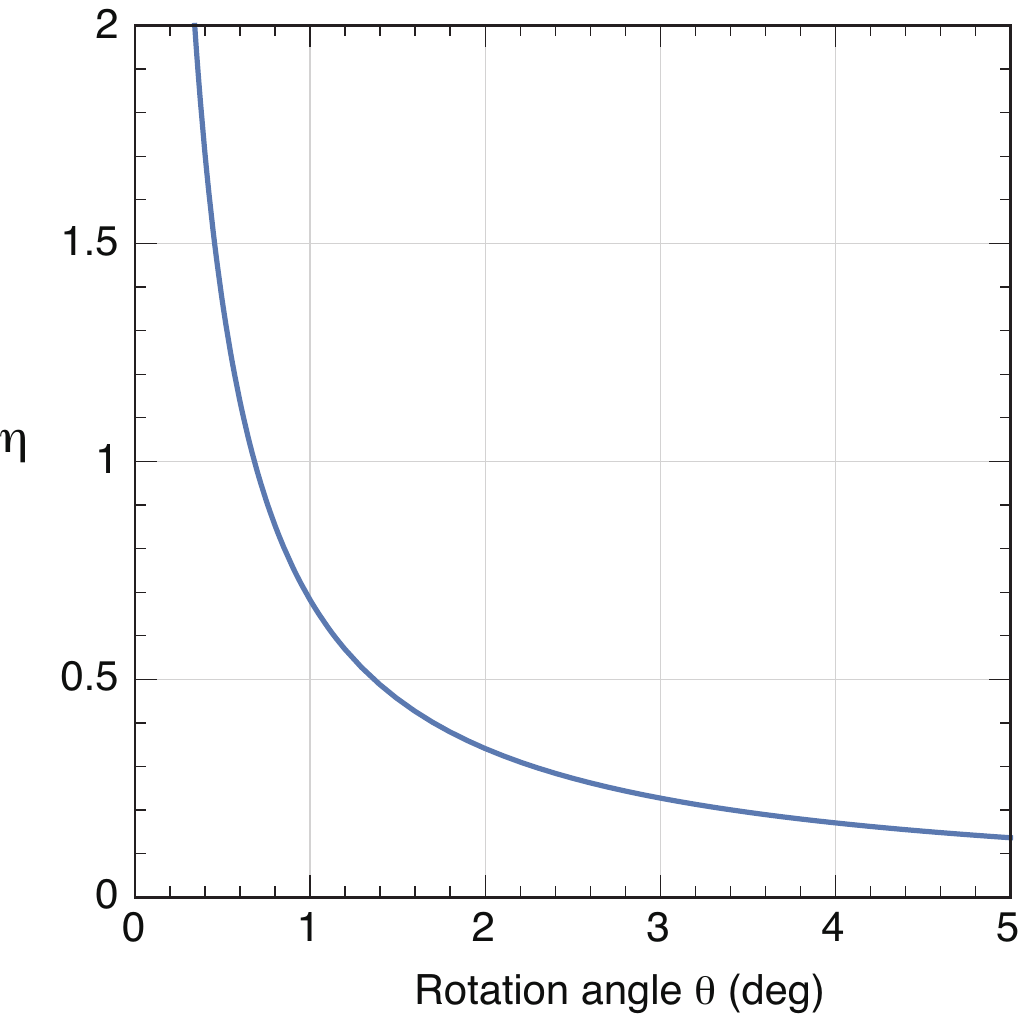}
\end{center}
\caption{Dimensionless parameter $\eta$ as a function of rotation angle $\theta$.}
\label{fig_eta_vs_theta}
\end{figure}

Eq. (\ref{Eq:Fourier}), (\ref{Eq:FqFunc}) and (\ref{Eq:E-LMatrix}) are 
a set of self-consistent equations.
Following Eq.\ (\ref{Eq:Eta}) in the 1D model, the number of the relavant harmonics in $\textbf{u}^{-}_{\textbf{q}}$ 
is roughly characterized by a dimensionless parameter
\begin{eqnarray}
\eta = \sqrt{\frac{V_0}{\lambda + \mu}} \,\frac{L_{\rm M}}{a}.
%\eta = \frac{V_0 L_{\rm M}^2}{(\lambda + \mu) a^2}.
\label{Eq:Eta_2D}
\end{eqnarray}
In TBG, we have two elastic constants $\lambda$ and $\mu$,
and it is ambiguous which should replace the position of $\kappa$ in Eq.\ (\ref{Eq:Eta}) for 1D model.
Here we adopt the simple sum $\lambda + \mu$ in Eq.\ (\ref{Eq:Eta_2D}).
Figure \ref{fig_eta_vs_theta} plots $\eta$ as a function of the rotation angle $\theta$.
The approximation with the lowest harmonics
(i.e., six $\textbf{q}$-points of $\pm\textbf{G}^{\rm M}_1,\pm\textbf{G}^{\rm M}_2,\pm\textbf{G}^{\rm M}_3$) is valid
when $\eta \ll 1$, or $\theta >\sim 2^\circ$.
The contribution of high frequency harmonics is not negligible when $\eta$ is of the order of 1.
			
\begin{figure*}
\begin{center}
\leavevmode\includegraphics [width=0.7\hsize] {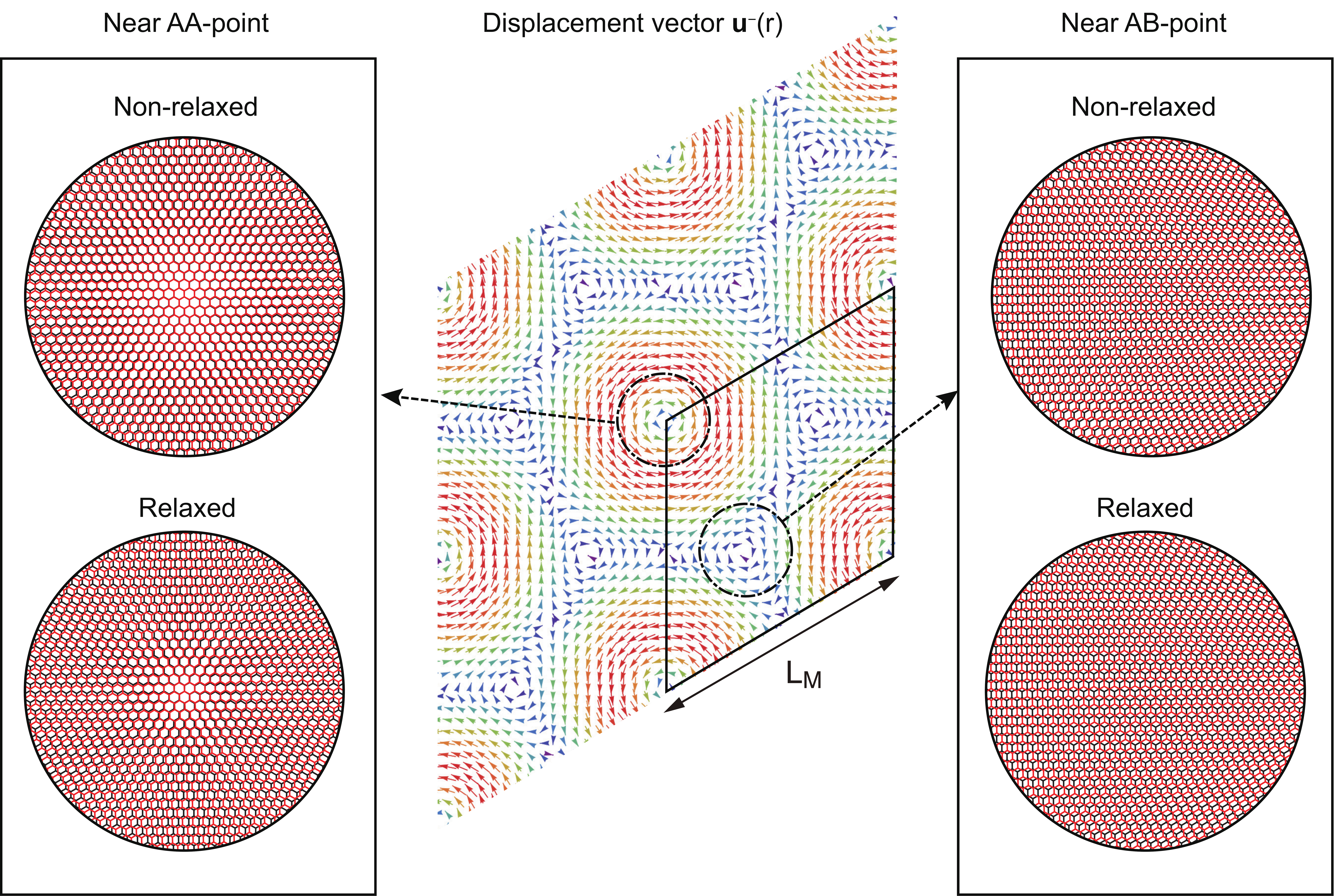}
\end{center}
\caption{
Center panel: Distribution of the displacement vector $\textbf{u}^{-} (\textbf{r})$
in the TBG of  $\theta= 1.05^{\circ}$. Left (right) side-panels: 
local atomic structure near AA (AB) stacked point before and after the relaxation. 
The small dashed circles in the center panel indicate the areas where the local structure is sampled.}
\label{fig_lattice_1.05}
\end{figure*}

 \begin{figure*}
\begin{center}
\leavevmode\includegraphics[width=0.9\hsize]{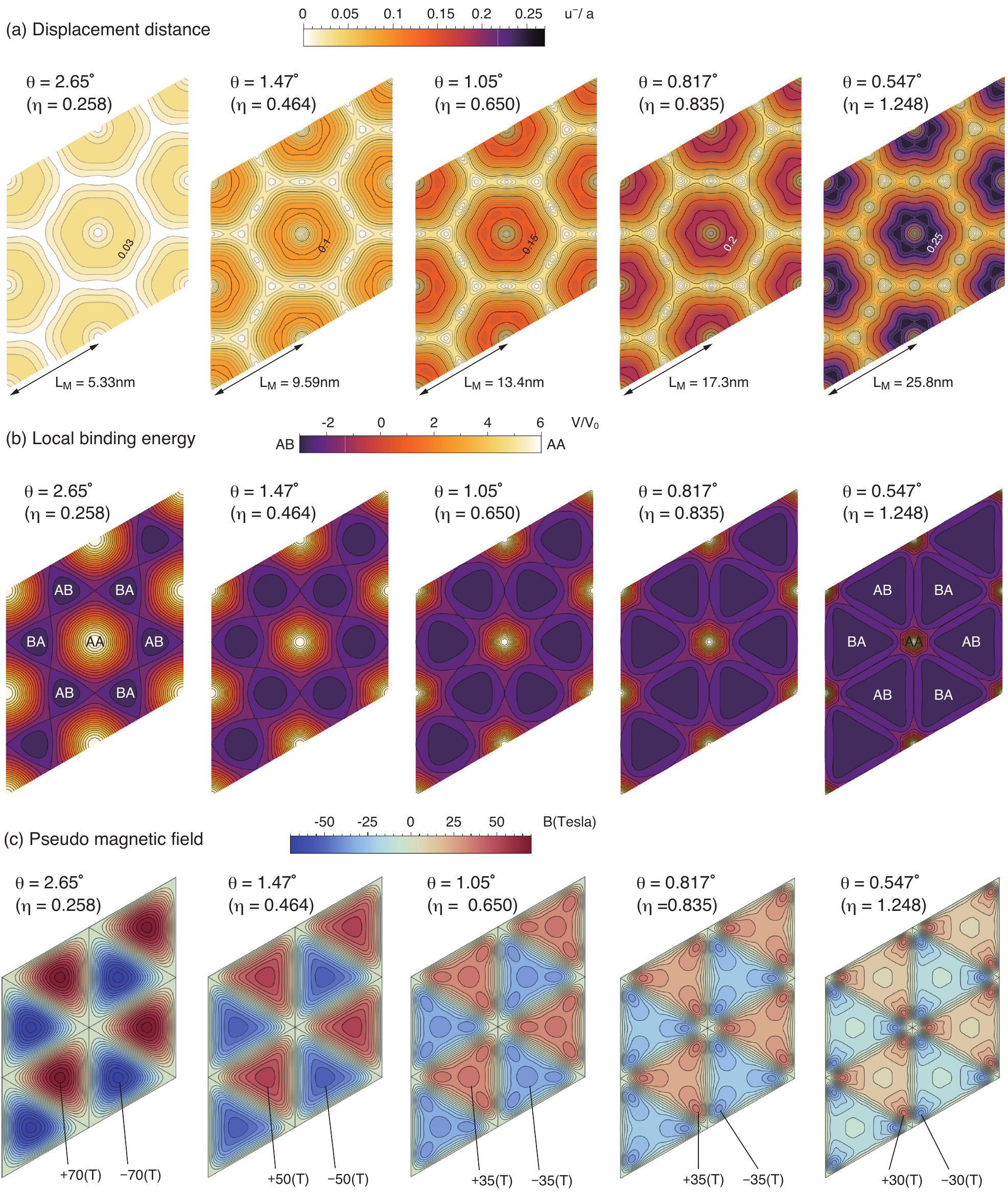}
\end{center}
\caption{Two-dimesional maps for
(a)  absolute value of the displacement vector $\textbf{u}^{-} (\textbf{r})$,
(b) the local binding energy $V[\bm{\delta} (\textbf{r})]$.
and (c) strain-induced pseudo magnetic field $B_{\rm eff} (\textbf{r})$,
calculated for TBGs with various rotation angles.}
\label{fig_u_tbg}
\end{figure*}

We numerically solve the self-consistent equation for several TBGs 
by numerical iterations with sufficiently large cut-off in $q$-space.
Figure \ref{fig_lattice_1.05} presents an example of calculated result for lattice relaxation in $\theta=1.05^\circ$,
where the central panel plots the displacement vector $\textbf{u}^{-} (\textbf{r})$ as a function of position,
and the left (right) panels show the local atomic structure near AA (AB) stacked point before and after the relaxation.
The actual displacement on each layer is 
given by $\textbf{u}^{(1)} = -\textbf{u}^{-}/2$ and $\textbf{u}^{(2)} = \textbf{u}^{-}/2$.
We actually observe that the  $\textbf{u}^{-}$ rotates around the center of AA region, and 
as a result, the AA region is significantly shrunk while AB region is expanded.

Figure \ \ref{fig_u_tbg} summarizes the results for TBGs from 
$\theta= 2.65^\circ$ down to $0.547^\circ$.
Here the panels in (a) show the absolute value of the displacement vector $\textbf{u}^{-} (\textbf{r})$
as a function of position.
The distribution of $\textbf{u}^{-} (\textbf{r})$ on two-dimensional place looks all similar among all the cases,
where it takes the maximum on a ring-like region near the AA spot.
On the other hand,  its magnitude strongly depends on $\theta$,
where the $|\textbf{u}^{-}|$ is much smaller than the atomic scale $a$
in $\eta \ll 1$, while it eventually becomes comparable when $\eta$ is of the order of 1.  
Figure \ref{fig_u_tbg}(b) presents the corresponding plots for
the local binding energy $V[\bm{\delta} (\textbf{r})]$.
When $\textbf{u}^{-} (\textbf{r})$ is much smaller than $a$, as in $\theta = 2.65^\circ$,
the potential profile is approximately given by $V[\bm{\delta}_0 (\textbf{r})]$,
which is essentially a sum of three plain waves.
In decreasing $\theta$, the spots of $AA$-regions shrink and AB and BA regions eventually dominate.
The result looks especially dramatic in small angles less than $1^{\circ}$,
where the relaxed lattices clearly exhibits a triangular domain pattern of AB and BA regions.
Similar to Eq.\ (\ref{eq_domain_width}) for the 1D model,
the characteristic width of the domain boundary is given by
\begin{align}
w_{\rm d} \approx \frac{a}{4}\sqrt{\frac{\lambda+\mu}{V_0}} \approx  5.2{\rm nm}.
\label{eq_domain_width_2d}
\end{align}	
Indeed it roughly agrees with the typical scale of the AB/BA domain wall in Fig.\ \ref{fig_u_tbg}(b).
It is also consistent with the experimental observation of the shear boundary,
which estimates the averaged width about 6 nm.\cite{alden2013strain}

\section{Band structure}
\label{sec_band}

To calculate the energy band structures in the presence of the lattice strain,
we use the tight-binding method. The Hamiltonian is written as
\begin{equation}
H = -\sum_{i,j} t(\textbf{R}_i -\textbf{R}_j) |\textbf{R}_i \rangle \langle \textbf{R}_j| + \text{h.c.}
\end{equation}
where $\textbf{R}_i$ is the atomic coordinate, $ |\textbf{R}_i \rangle $ is the wavefunction at site $i$, 
and $t(\textbf{R}_i -\textbf{R}_j)$ is the transfer integral between atom $i$ and $j$. 
We adopt the Slater-Koster type formula for the transfer integral,
 \cite{slater1954simplified}
\begin{equation}
-t (\textbf{d}) = V_{pp\pi} (d) \left[1 - \left(\frac{\textbf{d} \cdot \textbf{e}_z}{d}\right)^2 \right] + 
 V_{pp\sigma} (d)  \left(\frac{\textbf{d} \cdot \textbf{e}_z}{d}\right)^2
 \label{eq_t}
\end{equation}
\begin{align}
& V_{pp\pi} (d) = V_{pp\pi}^0  \exp \left(- \frac{d-a_0}{r_0} \right), \\
& V_{pp\sigma} (d)  =V_{pp\sigma}^0  \exp \left(- \frac{d-d_0}{r_0} \right)
\end{align}
where $\textbf{d} = \textbf{R}_i - \textbf{R}_j$ is the distance between two atoms. $\textbf{e}_z$ is the unit vector on $z$ axis. 
$V_{pp\pi}^0 \approx -2.7$eV is the transfer integrals between nearest-neighbor atoms of monolayer graphene which 
are located at distance $a_0 = a/\sqrt{3} \approx 0.142$nm. $V_{pp\sigma}^0 \approx 0.48 $eV is the transfer integral 
between two nearest-vertically aligned atoms. $d_0 \approx 0.334$nm is the interlayer spacing. 
The decay length $r_0$ of transfer integral is chosen at $0.184 a$
so that the next nearest intralayer coupling becomes $0.1 V_{pp\pi}^0$.
\cite{uryu2004electronic,trambly2010localization} 
At $d > \sqrt{3}a$, the transfer integral is very small and negligible.

\begin{figure*}
\begin{center}
\leavevmode\includegraphics[width=0.8\hsize]{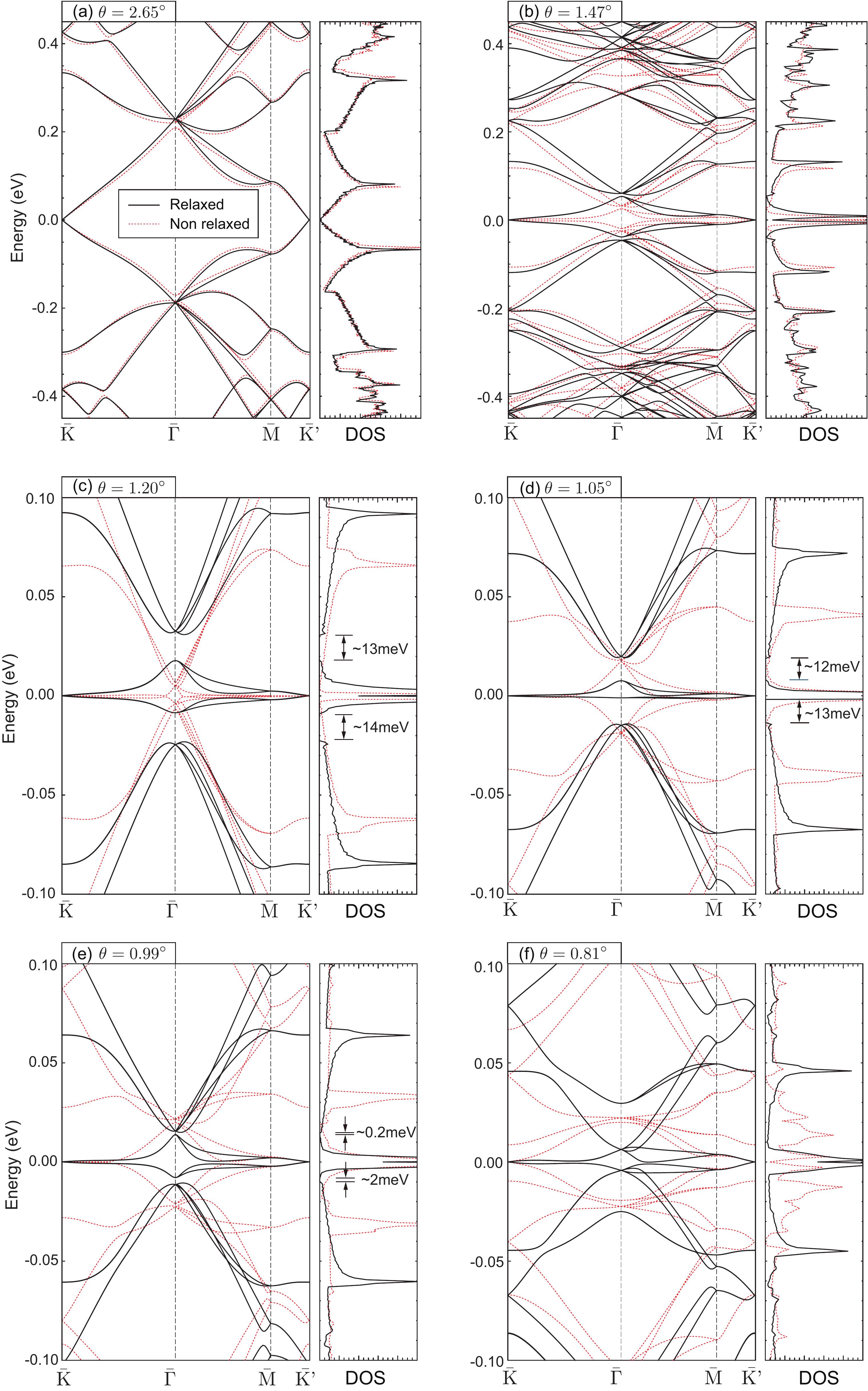}
\caption{Band structure and density of state  
of relaxed (black solid lines) and non-relaxed (red dashed lines) TBGs at
various rotaion angles. The energy gap is indicated by a pair of arrows in the DOS panel.
}
\label{fig:Band_and_DOS}
\end{center}
\end{figure*}

\begin{figure}
\begin{center}
\leavevmode\includegraphics[width=0.7\hsize]{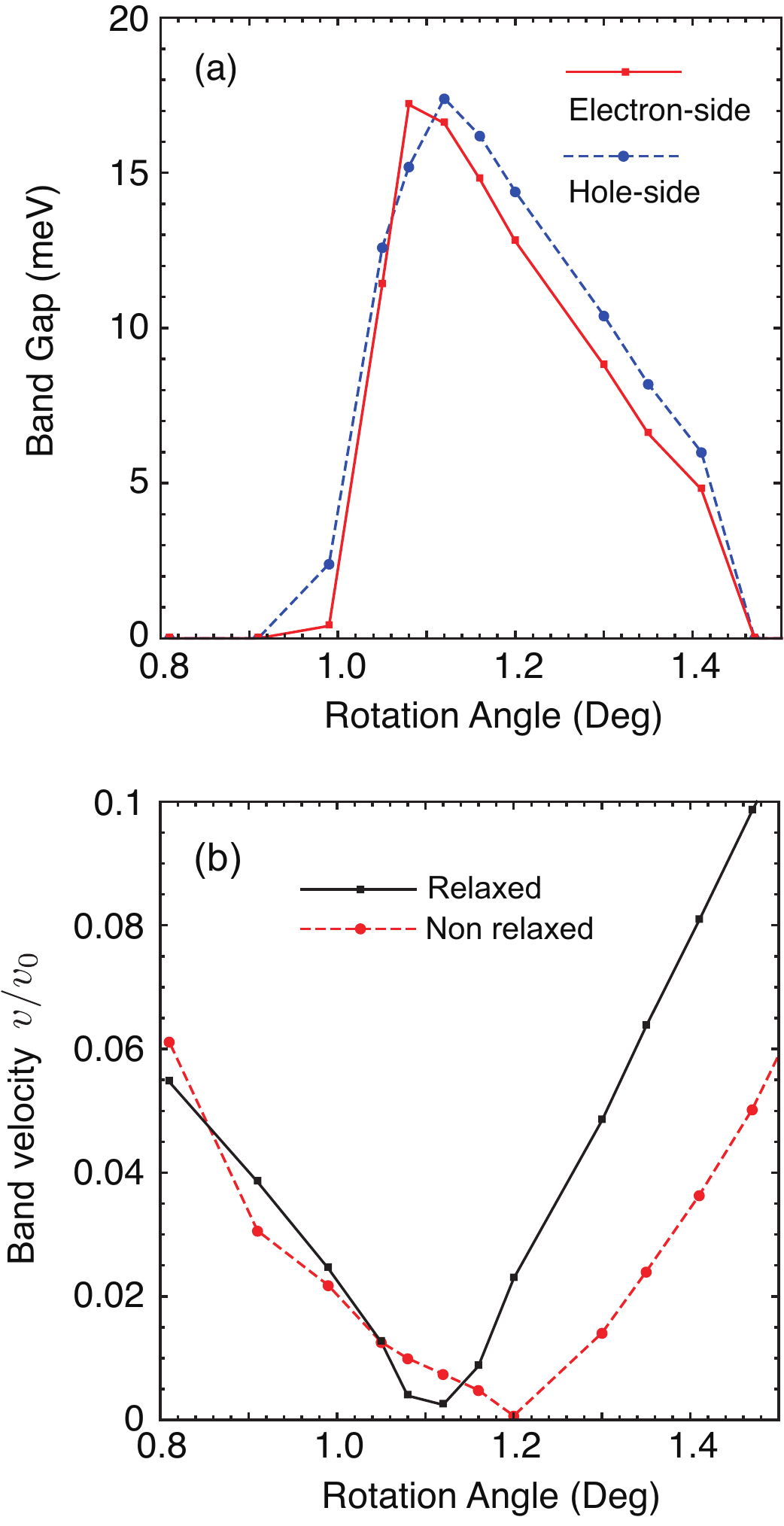}
\caption{(a) Band gap between the lowest band and the first excited bands in the electron side (red, solid line)
and in the hole side (blue, dashed), in relaxed TBGs against the rotation angle.
 (b) Band velocity at $K$-point as a function of $\theta$ for relaxed (black, solid) and non-relaxed (red, dashed) TBGs}
\label{fig:Velo_Bandgap}
\end{center}
\end{figure}	

Using the optimized structure obtained in the last section,
we specify the lattice position of each single atom in the relaxed TBG,
construct the tight-binding Hamiltonian, and calculate the energy bands.
Figure \ref{fig:Band_and_DOS} compares the electronic band structure
of relaxed (black solid lines) and non-relaxed (red dashed lines) TBGs at several rotation angles. 
The horizontal axis are labeled by the symmetric points of the Brillouin zone
for the moir\'{e} superlattice [Fig.\ \ref{fig:Lattice_Constant_BZ}], and it scales in proportion to $2\pi/L_{\rm M}$. 
At $\theta = 2.65^{\circ}$, we only see a minor difference
in accordance with the small change in the lattice structure in Fig.\ \ref{fig_u_tbg}(a).
A significant deviation is observed below $2^{\circ}$.
The most notable change from the non-relaxed case is that 
a band gap opens between the lowest subband near the Dirac point and the first excited subband
both in the electron side and the hole side.
Fig. \ref{fig:Velo_Bandgap} (a) shows the size of gap versus rotation angle $\theta$.
The gap is observed in TBGs of  $ 1^{\circ} \lsim \theta \lsim 1.5^{\circ}$,
and the maximum energy width is about 18meV. 

The lattice strain also strongly modifies the band velocity.
Figure \ref{fig:Velo_Bandgap} (b) plots the band velocity at the Dirac point
as a function of $\theta$ for relaxed and non-relaxed cases.
In both cases, the central band at the Dirac point is gradually flattened in decreasing $\theta$,
and the Fermi velocity vanishes at a certain angle \cite{TramblyDeLaissardiere2012confined,Bistritzer2011magicangle}.
We find that the band flattening is a little slower in the relaxed case,
i.e at the same angle, the band width is larger in the relaxed TBG than in non-relaxed counterpart,
so that the critical angle for the vanishing velocity is shifted to the lower rotation angle in the relaxed TBG.

The lattice relaxation affects the electronic structure
in two different ways,
by a change of interlayer Hamiltonian associated with the modified moir\'{e} pattern,
and also by a change of the intralayer Hamiltonian through distortion of the lattice.
The latter is known to be described by the pseudo magnetic field in the effective mass Dirac Hamiltonian.\cite{suzuura2002phonons,pereira2009strain,guinea2010energy}
The vector potential for the pseudo-magnetic field on layer $l (=1,2)$ is given by \cite{suzuura2002phonons,pereira2009strain,guinea2010energy}
\begin{eqnarray}
A^{(l)}_x &=& \frac{3}{4}\frac{\beta\gamma_0}{ev}[u^{(l)}_{xx}-u^{(l)}_{yy}], \\
A^{(l)}_y &=&  \frac{3}{4}\frac{\beta\gamma_0}{ev} [-2u^{(l)}_{xy}].
\label{eq_A}
\end{eqnarray}
where 
$\gamma_0=t(a_0)$ 
is the nearest neighbor transfer energy of intrinsic graphene, 
$v = (\sqrt{3}/2)a \gamma_0$ is the band velocity of the Dirac cone, and
\begin{eqnarray}
\beta = -\frac{d\ln t(d) }{d\ln d}\Bigr|_{d=a_0}.
\end{eqnarray}
In the present model Eq.\  (\ref{eq_t}), we have $\beta = a_0/r_0 \approx 3.14$.
The pseudo magnetic field is given by $B^{(l)}_{\rm eff} = [\nabla\times\Vec{A}^{(l)}]_z$.

Figure \ref{fig_u_tbg}(c) shows the distribution of 
or the pseudo magnetic field on the layer 1 for several TBG's. 
The field direction is opposite between layer 1 and 2,
because $\textbf{u}^{(1)} = -\textbf{u}^{(2)}$.
We observe a triangular pattern with positive and negative field domains,
which are centered at the AB and BA stacking regions, respectively.
The field amplitude is huge, but 
it does not necessarily results in a strong effect on the electronic structure,
since it is rapidly oscillating in space with nano-meter scale.
The pseudo-magnetic field enters in the Hamiltonian as a form of $evA$ with the pseudo vector potential $A$.
%so the scale which really matters is not $B$-field amplitude itself but the energy scale of $evA$.
When the field spatially modulates with the wave length $L_{\rm M}$, the associated matrix element opens 
a band gap 
at the energy $E \sim \hbar v / L_{\rm M}$ measured from the Dirac point.
Therefore, the effect of the pseudo field significantly affects the band structure
when $evA >\sim \hbar v / L_{\rm M}$, while otherwise it is just perturbative.
Now the scale of $evA$ is roughly estimated as
\begin{equation}
evA \sim \beta\gamma_0 u_{ij} \sim 2 \beta\gamma_0 \frac{u}{a} \sin\frac{\theta}{2},
\end{equation}
where the strain tensor $u_{ij}$ is estimated about $u/L_{\rm M} = 2(u/a) \sin(\theta/2)$,
considering the displacement field $\Vec{u}$ is modulating with the moire wavelength $\sim 1/L_{\rm M}$.
The typical scale of $u/a$ can be read from Fig.\ \ref{fig_u_tbg}(a).
For $\theta= 2.65^\circ$, for example, $u/a \sim $ 0.03 gives $evA\sim$ 10meV,
and it is much smaller than  $\hbar v / L_{\rm M} \sim $  180meV.
So the effect of $evA$ is perturbative, and this is consistent with a small change in the band structure 
observed in Fig.\  \ref{fig:Band_and_DOS} (a).
For $\theta= 1.05^\circ$, on the other hand, $evA\sim $ 30meV 
is comparable with $\hbar v / L_{\rm M}\sim$ 50meV,
so the pseudo field plays a significant role in the modification of the low-energy bands.

\section{Conclusion}
\label{sec_conclusion}

We have developped the effective theory to calculate the spontaneous relaxation in TBG, 
and studied the atomic and electronic structures. 
In rotation angle larger than $2^{\circ}$, the lattice is hardly deformed and so the effect on 
the electronic structure is minimal, while in smaller rotation angle below $2^{\circ}$, 
the lattice is significantly modified to form AB/BA triangular domain structure.      
The electronic band structure is then strongly modified
where a band gap up to 20 meV opens above and below the lowest band.
The lattice deformation also significantly relaxes the band flattening observed in non-relaxed case,
and it lowers the critical angle at which the Fermi velocity vanishes.

Actually a recent experiment observed 
an insulating gap about 50 meV at the superlattice subband edges
in TBG with $\theta \approx 1.8^\circ$. \cite{cao2016superlattice} 
This seems qualitatively consistent with the present result, although 
1.8$^\circ$ is out of the gap-opening range in our model calculation,
and also 50 meV is a bit too large compared to the typical gap width obtained here.
As we see in the present work, however, the lattice relxation and the electronic structure
sensitively depend on the parameter $\eta$,
and it might be possible that the real system could have a greater interlayer interaction,
allowing a greater gap and a wider range in the rotation angle for gap opening.
It is also conceivable that the gap could be enhanced
when the Fermi energy is right at the superlattice gap position,
while the doping effect is not considered in the present study.
We leave the further quantitative arguments for a future problem.

The present model takes account of only the in-plane components of the lattice distortion,
as it is aimed to describe the domain formation within the simplest theoretical framework.
Inclusion of the out-of-plane motion is known to give rise to a corrugation in the perpendicular direction,
\cite{uchida2014atomic,van2015relaxation,dai2016twisted, jain2016structure}
where the interlayer spacing modulates by 10\%.
In a small angle TBG less than 2$^\circ$, in particular, the detailed computational study
that the the interlayer spacing is largest only near AA spot while it is almost flat otherwise. \cite{van2015relaxation,dai2016twisted}
The corrugation is small even at the AB/BA domain boundary, 
presumably because it is a shear boundary with no tensile strain,
and also the optimized interlayer spacing does not strongly depend on the stacking structure around there.
Therefore, we expect that the corrugation effect on the electronic structure exclusively comes from
AA spots, where the interlayer distance change should reduce the interlayer electronic coupling by a few 10\%.
Since the system is dominated by AB/BA regions, the change of the electronic band structure
is expected to be minor compared to the change caused by AB/BA domain formation itself.

\section*{ACKNOWLEDGMENTS}  
The authors thank Pilkyung Moon for helpful discussions. 
This work was supported by JSPS KAKENHI Grants No. JP25107001, No. JP25107005 and JP15K21722.

\bibliography{TBG_With_Strain}

\end{document}